# Fast Control Latency Uncertainty Elimination for the BESIII ETOF Upgrade[*]

Yun Wang(汪昀)[1, 2], Ping Cao (曹平)[1, 2;1)], Shu-bin Liu (刘树彬)[1, 3], Qi An (安琪)[1, 2]

[1]*State Key Laboratory of Particle Detection and Electronics, University of Science and Technology of China, Hefei, 230026, China*
[2]*School of Nuclear Science and Technology, University of Science and Technology of China, Hefei 230027, China*
[3]*Department of Modern Physics, University of Science and Technology of China, Hefei 230027, China*

**Abstract:** A new fanning topology is proposed to precisely fan out fast control signals in the Beijing Spectrometer (BES III) end-cap time-of-flight (ETOF) electronics. However, uncertainty in transfer latency is introduced by the new fanning channel, which will degrade the precision of fast control. In this paper, latency uncertainty elimination for the BESIII ETOF upgrade is introduced. The latency uncertainty is determined by a Time-Digital-Converter (TDC) embedded in a Field-Programmable Gate Array (FPGA) and is eliminated by re-capturing at synchronous and determinate time. Compared with the existing method of Barrel-cap TOF (BTOF), it has advantages of flexible structure, easy calibration and good adaptability. Field tests on the BES III ETOF system show that this method effectively eliminates transfer latency uncertainty.

## 1 Introduction

The Beijing Electron Positron Collider (BEPC) and the Beijing Spectrometer (BES) were upgraded to BEPC II and BES III respectively in the summer of 2008. To further improve the overall time resolution of the particle identification (PID), the end-cap time-of-flight system (ETOF) is being upgraded with newly developed MRPC detectors [1,2,3]. After upgrade, the total time resolution will be improved from 138 ps to about 80 ps [4]. To achieve this specification, new readout electronics for the ETOF should be developed.

In the existing BESIII TOF system, there are 2 VME crates, each of which has a fast control module (FCTL) inside receiving signals from the fast control system (FCS). For the new ETOF electronics, two extra VME crates are added to adapt to the large number of MRPC channels. However, there are only two critical and calibrated channels between the FCS and the TOF sub-system. A new topology needs to be proposed for transmitting fast control signals to the new ETOF electronics.

There are several different methods to transmit fast control signals in particle experiments. In the Babar experiment, fast control signals are transmitted through electric cables without being encoded [5]. There is no transmission latency uncertainty in this experiment. In the LHC experiments, fast control signals are transmitted through dedicated ASIC chips designed by CERN [6]. The transmission latency uncertainty is eliminated by special techniques of encoding and synchronizing code [7]. In BESIII, fast control signals are distributed with a serializing and de-serializing (SerDes) technique [8]. Fast control signals are encoded from 8-bit to 10-bit data with timing information and then serialized into bit stream for output at the transmitting side. However, at the receiver side, there might be 10 discrete phases when recovering clock from the bit stream. These 10 phases lead to latency uncertainty in distributing fast control signals. In the BESIII FCS design, the fast control data is re-synchronized at a determinate time to avoid the area of uncertainty [8,9]. This method succeeds only when the lengths of the distributing channels are equal. However, two extra VME crates change the distributing topology for ETOF fast control signals, which leads to the failure of the existing method. In this paper, a new method is presented. This method uses a TDC to calculate the uncertainty area automatically. It is easy to use and can be performed automatically.

## 2 Fast Control Latency Uncertainty in ETOF upgrade

### 2.1 Structure after upgrade

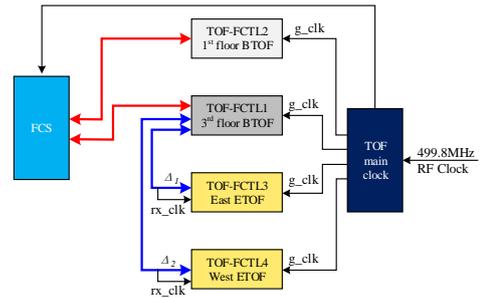

Fig. 1 Structure for fast control signal fanning out after ETOF upgrade

There are several sub-systems such as the TOF, muon detector, MDC and EMC connected to the BESIII FCS [8]. The FCS resides in the BES III trigger system. Inside the FCS, there are five fiber channels dedicated to fan fast control signals (L1 trigger signal, fiber reset signal etc.) out to one sub-detector system. For the TOF fast control sub-system, there are 4 electronic modules (TOF-FCTL1~4 in Fig. 1) for receiving signals distributed from the FCS. As shown in Fig. 1, there are two for the ETOF and two for the BTOF. There are two clocks on each ETOF fast control module, one being the global clock (g_clk) distributed from the TOF main clock, and the other being the rx_clk recovered from the bit stream (blue lines in Fig. 1) sent from the BTOF. The phase difference between these two clocks is noted as $\Delta$.

---

[*] Funded by CAS maintenance project for major scientific and technological infrastructure (IHEP-SW-953/2013)
1) E-mail: cping@ustc.edu.cn

In order to avoid modifications of the existing FCS structure, there are only two fiber channels (red lines in Fig. 1) left for TOF-FCTL. In the ETOF electronics, the two new VME crates need two additional fiber channels. In Fig. 1, to distribute fast control signals to all the TOF-FCTL modules, the module residing in the 3rd floor BTOF crate acts as a router. It receives fast control signals from the FCS then fans them out to each ETOF module. Two extra channels (blue lines in Fig. 1) are added in the new distribution topology.

## 2.2 Uncertainty introduced by SerDes

The existing FCS uses TLK1501 [10] to distribute fast control signals to all sub-systems. The same technique is used in the ETOF module design for compatibility purposes. The basic principle for using the SerDes technique is that the transmitter sends encoded data according to a reference clock, while the receiver acquires data according to the clock recovered from the transmitted bit stream, so the acquired data is aligned to the recovered clock. In Fig. 2, the yellow signal denotes the transmitting reference clock, while the green one denotes the recovered clock. Unfortunately, each time the SerDes powers up, there is uncertainty in the phase of the recovered clock as shown in Fig. 2, where the oscilloscope runs in persistence mode. This uncertainty is not a problem for traditional communication applications, where the correctness of data is the key point. However, for the BESIII FCS distributing fast control signals, the most critical point is the time of the fast control signal being recognized.

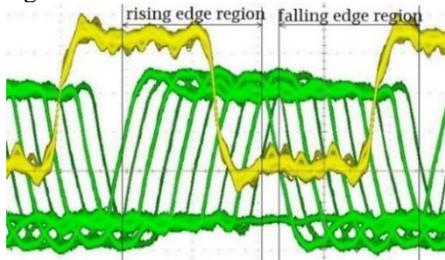

Fig. 2 Phase uncertainty in the recovered clock.

For BESIII, the clock period is 24 ns. The recovery clock has ten phases, which are uniformly spaced in the 10.2 ns range. As the work clock of the TOF-FCTL is the global clock, the recovery data must be re-synchronized by the global clock. Because the phase difference Δ might take any value, the rising edge of the global clock might lie in the rising edge region of the recovery clock. The transfer latency will have one cycle (24 ns) difference between the first recovery clock phase and the last one.

In the existing BTOF module, the encoded fast control data capturing time is postponed to avoid the unstable area (responding to the uncertain recovered clock) in the data bus according to the global synchronous clock, not the recovered clock [8,9]. The transmitted data is re-captured by the global synchronous clock which is shifted by a DLL with configurable delay ability. This can succeed in the existing TOF sub-system based on two facts. Firstly, there is a global precise clock network everywhere in the BESIII electronics. Secondly, there are only two identical distribution channels (red lines in Fig. 1) from the FCS to the TOF sub-system, which means the acquisition time can be easily calibrated and adjusted. During calibration, the FCS and TOF are placed in the same place and both are set into debugging mode. The FCS continuously sends trigger data to the TOF. At the TOF side, it is convenient to adjust the delay value by comparing the transmitted and received trigger signal with an oscilloscope.

Unfortunately the existing method cannot be applied in the ETOF upgrade, as the manual field test is very hard. The ETOF electronics should have the ability to eliminate latency uncertainty automatically in the spectrometer field.

## 3 Uncertainty Elimination

### 3.1 Description of principle

Because of the uncertainty in the recovered clock (rx_clk) at the receiver side, there is an unstable period in the fast control data which is recovered according to the rx_clk. This unstable period is 1.2 ns less than a half cycle as area P or N shown in Fig. 3. The rx_clk has the same frequency as the global synchronous clock, so if the recovered data is sampled at both edges of a reference clock (shifted from the TOF global clock, abbreviated as ref_clk), at least one value is reliable. Figure 3 shows the timing graph where the rx_data (16-bit width) is aligned with rx_clk.

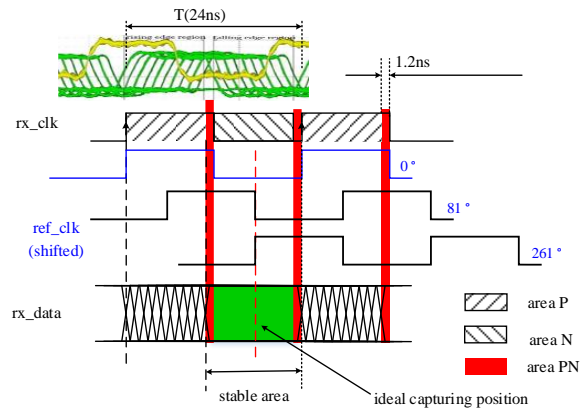

Fig. 3 Timing graph for elimination principle

One of ref_clk's edges (leading or trailing) must be in the stable area of rx_data. To improve the reliability of data capture, the middle position (red dashed line in Fig. 3) is the ideal capturing position, which makes the capture margin large enough. Furthermore, this position is determinate and does not change whenever SerDes powers up. Supposing there is no skew between rx_clk and ref_clk ($\Delta=0$ in Fig. 1), if the position of rx_clk is fixed and that of ref_clk is then checked, there will be 10 randomly distributed phases, among which both 81° (falling edge) and 261° (rising edge) correspond to this ideal position (bold black lines shown in Fig. 3).

The problem left is to obtain the clock with phase 81° (or 261°). This finding process can be considered as checking the relative position between rx_clk and ref_clk equivalently. There is no need to exactly distinguish each position among these 10 phases. The principle is to make the real capturing position move as close to the ideal position as possible. So the relative phase difference can be measured by a TDC with bin size of 3 ns. One clock cycle ($T$) can be divided into 8 parts with initial phase value from $\phi_0$ to $\phi_7$, where $\phi_i=i*45°$. If the TDC's dynamic range is only one cycle (T), it will generate 8 data points (0~7) corresponding to each phase respectively (Fig. 4).

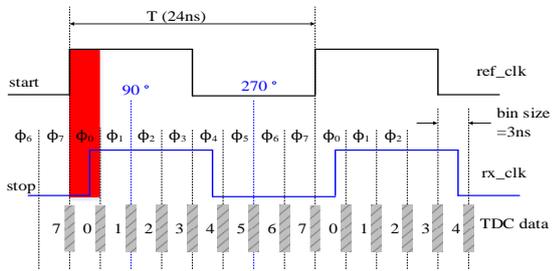

Fig. 4 TDC data and clock phase

In Fig. 4, the rising edge of ref_clk is used to start the TDC counting, while the rising edge of rx_clk is used to stop it. If the phase difference is less than one bin size (red block in Fig. 4), the TDC will give an output value of 0. So according to the output code from the TDC, it is easy to know the relative position between rx_clk and ref_clk. However, because of the existence of clock jitter or bit unmatch of rx_data (16-bit width), the TDC may generate two values randomly during the critical period of bin transition (gray block in Fig. 4). For the case of 0 phase difference, TDC can generate 0 or 7, while for the case of 90°, the data is 1 or 2.

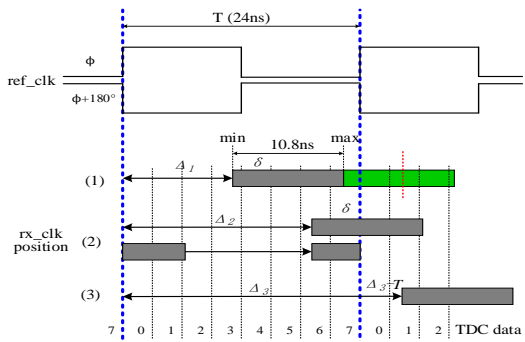

Fig. 5 Algorithm for finding rx_data re-capture position

Because of the existence of uncertainty in the rx_clk phase, the phase difference between ref_clk and rx_clk can be considered as follows:
$$\Delta\phi=\Delta+\delta$$
$\Delta$ denotes the difference caused by the signal distribution channel (blue lines in Fig. 1), and $\delta$ denotes the uncertainty caused by SerDes.

In Fig. 5, dark grey blocks denote the uncertain area where the rising edge of rx_clk occurs. The width of this area is 10.8 ns. The center of the attached area (green block in Fig. 5) is the ideal position for data capture. There are three possibilities for phase difference ($\Delta$) discussed as follows:

**(1) $0 \leq \Delta < T/2+1.2$**

The position of the rx_clk rising edge swings in one period $T$ as shown in Fig. 5. The TDC generates the following data:
{0, 1, 2, 3, 4} when $\Delta$ is digitized as 0;
{1, 2, 3, 4, 5} when $\Delta$ is digitized as 1;
{2, 3, 4, 5, 6} when $\Delta$ is digitized as 2;
{3, 4, 5, 6, 7} when $\Delta$ is digitized as 3.

**(2) $T/2+1.2 \leq \Delta < T$**

The position of the rx_clk rising edge overflows outside one period T as shown in Fig. 5. The TDC generates the following data:
{4, 5, 6, 7, 8}, when $\Delta$ is digitized as 4;
{5, 6, 7, 8, 9}, when $\Delta$ is digitized as 5;
{6, 7, 8, 9, 10}, when $\Delta$ is digitized as 6;
{7, 8, 9, 10, 11}, when $\Delta$ is digitized as 7.

Since the TDC dynamic range is only one cycle, the swing area (from TDC output data) will be folded back to the beginning of this cycle. So the data pair generated by the TDC is:
{4, 5, 6, 7, 0}, when $\Delta$ is digitized as 4;
{5, 6, 7, 0, 1}, when $\Delta$ is digitized as 5;
{6, 7, 0, 1, 2}, when $\Delta$ is digitized as 6;
{7, 0, 1, 2, 3}, when $\Delta$ is digitized as 7;

Compared with the first situation, the <*min*, *max*> value pair is definitely <0, 7> whichever value $\Delta$ is.

**(3) $\Delta > T$**

It can be classified into case 1 or 2 if the TDC counting start signal is switched to the next following rising edge of ref_clk until $(\Delta-n*T)<T$.

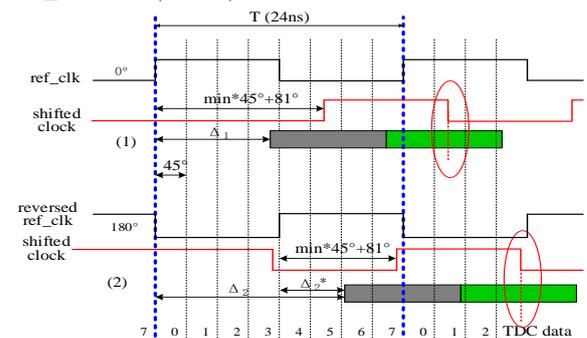

Fig. 6 Selection of capturing rx_data (green blocks denote the stable area for rx_data) at shifted clock (colored with red)

According to the above discussion, the algorithm for re-capturing rx_data at the receiving side is as follows:

(1) Receiver resets SerDes periodically to generate rx_clk with uncertain phase and set *rf*=0.

(2) Measure the phase difference ($\Delta<T$, normalized to one cycle) between rx_clk and ref_clk, and count the minimum and maximum value to form <*min*, *max*> pair.

(3) If <*min*, *max*> pair is not <0, 7>, shift ref_clk with phase ($min*45°+81°$) and then go to step (5).

(4) If <*min*, *max*> pair is <0, 7>, shift ref_clk with phase 180°, *rf*=1 and then go back to step (2).

(5) Capture rx_data with the shifted ref_clk (*rf*=0) or reversed ref_clk (*rf*=1) at its falling edge (red circle in Fig. 6) and then go to step (6).

(6) Finish.

### 3.2 Implementation

Figure 7 is the block diagram of the uncertainty elimination algorithm implemented in a FPGA. The recovered data (rxd with 16-bit width) is directly fed into a group of flip-flops inside one common logic array block (LAB) in the FPGA for data alignment. It is synchronized to rx_clk. Then the aligned data is sampled by two groups of flip-flops simultaneously, one working on phase $\phi$, the other working on $\phi+180°$ ($\phi$ is automatically obtained through the phase difference measuring flow, and the resolution of phase shifting is 200 ps). The two groups of flip-flops must be constrained together to ensure that the 16-bits aligned data are sampled under the same conditions. After being synthesized and implemented in the FPGA, timing analysis shows that the data skew of all the sample DFFs is only a few picoseconds. Finally, one of the sampled data points is selected as the stable output according to the above algorithm.

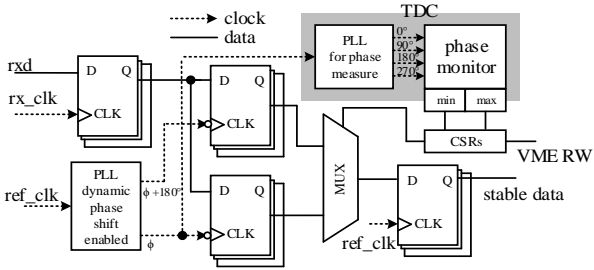

Fig. 7 Logic block diagram for uncertainty elimination

The sampling ref_clk phase can be adjusted by logic automatically. A multi-phase clock interpolation technique based TDC [11] implemented in FPGA is used to measure the phase differences of rx_clk and ref_clk. Fig. 8 is the schematic of the TDC. Because the phase difference is less than one clock period (24 ns), the TDC does not need a large dynamic range and the TDC structure can be further simplified. The TDC contains four parts as shown by the attached blocks in Fig. 8. The $\phi$ clock (also shown in Fig. 7) is the main clock for the TDC. The df_clk[3:0] is generated by doubling the frequency of the $\phi$ clock and shifting the phases from 0° to 270° respectively. The hit signal is generated by dividing the ref_clk by a factor of two. Each sample block generates two bits. All these bits are arranged and synchronized by the $\phi$ clock to generate the output *v[7:0]*. As the period of hit is twice as the $\phi$ clock, the TDC generates one output every two clock cycles. The phase monitor (shown in Fig. 7) calculates the maximum and minimum value of *v* continuously.

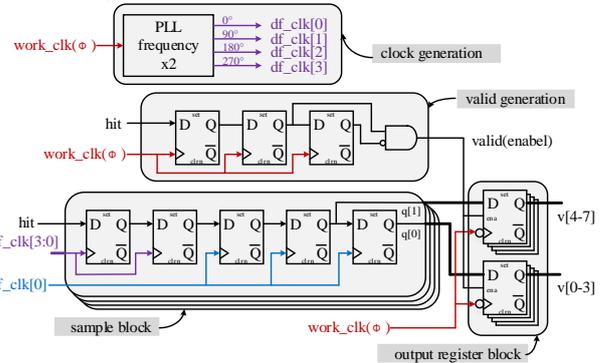

Fig. 8 The schematic of the TDC

Fig. 4 shows the characteristics of this TDC. The performance has been measured by an indirect way. The test is executed in the following steps:

(1) Enable the TLK1501 to generate stable rx_clk.

(2) Enable the TDC module to start measurement for one minute.

(3) Record the measured <*min, max*> pair. Disable the TDC module to clear the <*min, max*> value.

(4) Shift the $\phi$ clock by 200 ps. If the total phase shift reaches one cycle (24 ns i.e. 120 steps), then go to step (5). If not, go to step (2).

(5) Finish.

Fig. 9 shows the recorded data. Figure 9(a) and (c) show that each bin contains about 15 shift steps, which span 15*0.2 ns = 3 ns. The linearity is good. Figure 9(b) shows that at most one shift step has nonzero (max-min) value at each bin transition boundary. This indicates that the transition area (diagonal areas in Fig. 4) is in the range less than 2*200 ps = 400 ps including all contributions (such as clock jitter, hit jitter, DFF setup and hold time requirements). As the phase monitor does statistics on the output and the 400 ps transition region is far less than the 3 ns bin size, it will have no substantial effect on the methods proposed in section 3.1.

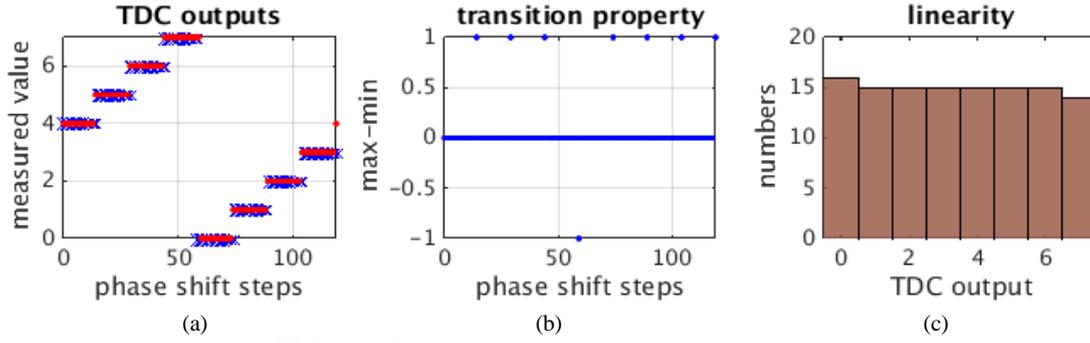

Fig. 9 The measured properties of the TDC. (a) is the output versus shift steps; (b) shows the differences between maximum value and minimum value; (c) shows the linearity of the TDC.

If the SerDes is triggered into a cycle of being enabled and disabled continuously, it will generate rx_clk with uncertain phase. The phase difference ($\Delta$) can be measured by the phase monitor.

## 4 Experiments & Verification

To verify this proposed method, an experiment was carried out. The test platform is shown in Fig. 10.

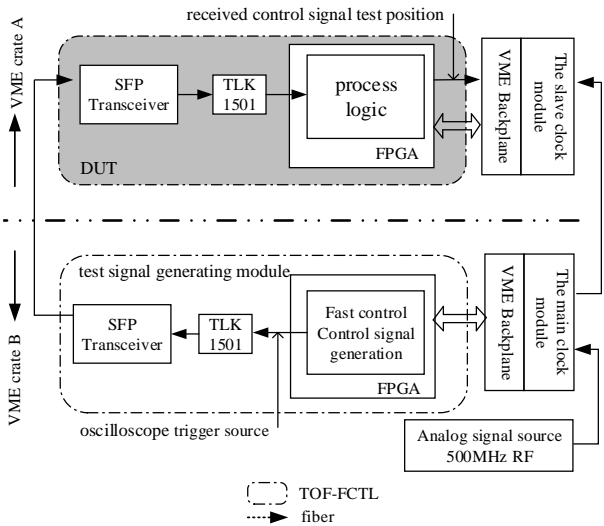

Fig. 10 Structure of the test platform

There are two VME crates (A and B). DUT (ETOF-FCTL) is settled in crate A, while a test signal generating module is settled in B. The test signal generator simulates the FCS behavior in the BESIII trigger system to fan out fast control signals. The generated test pulse width is 4 clock cycles, which is the same as the FCS. The delay between two successive test pulses is randomly controlled by a seven-order linear-feedback shift register (LFSR). In order to accelerate the test speed, the maximum delay is 128 clocks.

The test procedure was divided into two phases. In the first phase, we repeatedly reset the SerDes on the receiver side for calibration. In the second phase, we monitored the recovered fast control signal at the DUT side with an oscilloscope to check for uncertainty. This procedure was carried out in both laboratory and BESIII environments.

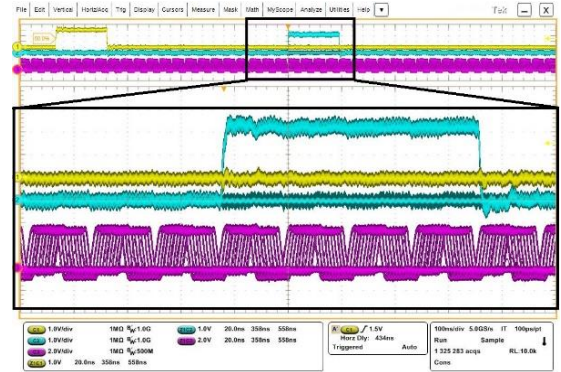

Fig. 11 Fast control signal uncertainty monitoring result (yellow line: transmitted signal, blue line: recovered signal, purple line: recovered clock; top half: global view, bottom half: enlarged view).

In Fig. 11, the transmitted test signal (yellow line) is used as the trigger source of the oscilloscope, while the recovered signal (blue line) at the DUT side is monitored in infinite presentation mode, which means the uncertainty between the transmitted and recovered signals can be captured by the oscilloscope.Fig. 11 shows that the latency between these two signals is fixed. The fast control signal uncertainty is eliminated. Fig. 11 wsa captured after this test platform had run for 8 hours with SerDes reset repeatedly, which means this proposed method has good stability.

Furthermore, the upgraded ETOF fast control module has been implemented with this method. Tests have also been carried out in the BESIII environment. Results show that the uncertainty is eliminated.The new modules have been installed into the BESIII upgrade ETOF VME crates (Fig. 10) and work well.

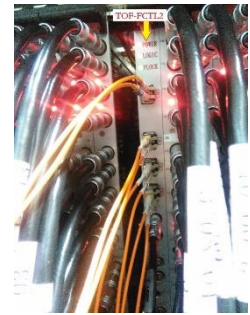

Fig. 12 Upgraded ETOF fast control module runs in BESIII

## 5 Conclusions

To further improve the overall time resolution of the BESIII TOF detector system, the ETOF system needs to be upgraded. In the upgrade architecture, there are 4 electronic modules which need to receive fast control signals transmitted from the BESIII FCS in the trigger system. However, there are only 2 transmitting channels between the FCS and the TOF system, which makes the transmitting topology differ from the existing one. Now, fast control signals are first directly transmitted to two electronic modules in the BTOF crates and then relayed to the ETOF modules by one BTOF module. The uncertainty in the fast control signal recovered from SerDes occurs in ETOF just as in the TOF before. However, there is a restriction that the BTOF modules should not be modified during the ETOF upgrade, and because the characterization of the new ETOF electronics differs from that of the BTOF, the existing uncertainty elimination method for the BTOF module no longer works for the ETOF. The ETOF modules need to be calibrated again with as little accelerator machine time as possible.

In this paper, a new method to eliminate the uncertainty is proposed. In this method, to guarantee the latency certainty, the transmitted fast control data is re-captured under the control of a clock shifted from the global synchronous clock, not the clock recovered from the SerDes bit stream. The exact shifting phase is determined by a TDC module embedded in an FPGA, which is designed and implemented with the data processing logic in the FPGA.

Compared with the existing method, it can automatically find a deterministic position with a good margin, which makes it better for the application of the ETOF upgrade without affecting any other existing sub-systems. The results of long time tests in the laboratory and in-situ in BESIII show that the method is correct and stable. Fast control latency uncertainty for the BESIII ETOF upgrade can be eliminated successfully.

**Acknowledgement**
*We gratefully acknowledge Dr. Hongliang Dai, Zhi Wu, Xiaolu Ji, Jingzhou Zhao, Xiaozhuang Wang and other members of BESIII ETOF upgrade group of IHEP for their earnest support and help during the beam test.*